\documentstyle[aps,epsf]{revtex}
 
\newcommand{\bq}{\begin{equation}}
\newcommand{\eq}{\end{equation}}
\newcommand{\bqn}{\begin{eqnarray}}
\newcommand{\eqn}{\end{eqnarray}}
\newcommand{\nb}{\nonumber}
\newcommand{\lb}{\label}
\begin{document}
 
\title{ 
Dynamics of Rotating Cylindrical Shells in General Relativity}
\author{Paulo R.C.T.
Pereira  \thanks{Email: terra@maxwell.on.br}} \address{Departamento de
Astrof\'{\i}sica, Observat\'orio Nacional~--~CNPq, 
Rua General Jos\'e Cristino 77, S\~ao Crist\'ov\~ao, 20921-400 
Rio de Janeiro~--~RJ, Brazil}
\author{Anzhong Wang  \thanks{Email: wang@dft.if.uerj.br}} 
\address{ Departamento de F\' {\i}sica Te\' orica,
Universidade do Estado do Rio de Janeiro,
Rua S\~ ao Francisco Xavier 524, Maracan\~a,
20550-013 Rio de Janeiro~--~RJ, Brazil}

\date{\today}

\maketitle

\begin{abstract}

 Cylindrical spacetimes with rotation are studied using the Newmann-Penrose
 formulas. By studying null geodesic deviations the physical meaning of
 each component of the Riemann tensor is given. These spacetimes
 are further extended to include rotating dynamic shells, and the general 
 expression of the surface energy-momentum tensor of the shells is given 
 in terms  of the discontinuation of the first derivatives 
 of the metric coefficients.
 As an application of the developed formulas, a stationary shell
 that generates
 the Lewis solutions, which represent the most general vacuum cylindrical 
 solutions of the Einstein field equations with rotation,
 is studied by assuming that the 
 spacetime inside the shell is flat. It is shown that the shell can satisfy
 all the energy conditions  by properly choosing
 the parameters appearing in the model, provided that
 $ 0 \le \sigma \le 1$, where $\sigma$ is related
 to the mass per unit length of the shell. 
\end{abstract}
 
%\noindent

{PACS numbers: 04.20Cv, 04.30.+x, 97.60.Sm, 97.60.Lf.}

%\newpage
 
%%%%%%%%%%%%%%%%%%%%%%%%%%%%%%%%%%%%%%%%%%%%%%%%%%%%%%%%%%%%%%%%%%%%%%%%%%%%%%%%
\section{Introduction}
 
 Gravitational collapse of a realistic body has been one of the most
thorny and important problems in Einstein's theory of
General Relativity. Due to the 
complexity of the Einstein field equations, the problem even in 
simple cases, such as, spacetimes with spherical symmetry, is still 
not well understood \cite{Josh1994}, and new phenomena keep
emerging \cite{Ch1993}. 
In 1991, Shapiro and Teukolsky \cite{ST1991} studied  numerically the
problem of a dust spheroid, and found that only the spheroid is compact
enough, a black hole can be formed.  Otherwise, the collapse most
likely ends with a naked singularity. Since then, the gravitational
collapse with non-spherical symmetry has attracted much attention. In
particular, by studying the collapse of a cylindrical shell that consists
of counter-rotating particles,
Apostolatos and Thorne (AT)  found that the rotation always
halts the collapse \cite{AT1992}. As a result, 
no naked singularities can be formed
on the symmetry axis.  However, in the AT work only the case
where the total angular momentum of the collapsing shell
is zero was considered. In more realistic
case, the spacetime has neither cylindrical symmetry nor zero angular
momentum. As a generalization of the AT work, in this paper we shall
consider the case where the total angular momentum 
is not zero, while still keep the requirement that the spacetime be
cylindrical.

Another motivation for us to consider rotating shell comes from recent 
study of the physical interpretation of the Levi-Civita vacuum solutions
\cite{WSS1997}, and the Levi-Civita solutions with cosmological
constant (LCC) \cite{SWPS1999}. These solutions have been known for a long
time \cite{Kramer1980}, but their physical properties were studied
extensively only very recently \cite{Bonnor1992}, and there are still several
open problems to be solved. By looking for some physical sources to the
LC solutions, we extended one of the two parameter that is related to,
but not equal to, the mass per unit length, from the range, $[0,
\frac{1}{4}]$, to the range,  
$[0, {1}]$ \cite{WSS1997}. However, beyond this range, the
physical meaning of the solutions is still not clear. Despite of the 
simplicity of the solutions, it was found
that they have very rich physical meaning. In particular,  
 they can give rise to black hole structures \cite{SWPS1999}.
Since the LC and LCC solutions are all static, it is very interesting to  
generalize these studies to the rotating case. 

As it can be seen from the discussions given below, when rotation is
 included, the problem is considerably complicated. This partially explains
why spacetimes with rotation are hardly studied (analytically). Thus, 
to start with, in the next section (Sec. II) we shall first study 
the main properties of cylindrical spacetimes 
with rotations, using the Newmann-Penrose (NP)
formulas \cite{NP1962}. One of the main reasons to use the NP formulas is that 
they give directly physical interpretation for each component of the Riemann
tensor. This is particularly useful when spacetimes contain 
gravitational waves. In Sec. III, the spacetimes are extended to include
the case where rotating matter shells appear.   To deal with the problem, one
usually uses Israel's method \cite{Israel1966}. However, we find that for the 
present problem Israel's method becomes very complicated and is very
difficult to be implemented. Instead, we shall follow Darmois
\cite{Da1927} and Lichnerwicz  \cite{Li1955} (see also Papapetrou and Hamoui
\cite{PH1968} and Taub \cite{Taub1980}). Although the two methods are 
essentially equivalent \cite{BV1981},   the latter is simpler, specially in
dealing with complicate boundaries like the present one. The disadvantage of
the latter is that it requires only one set of coordinates across 
the boundaries, while Israel's method does not. In this section, the physical
interpretation of the surface EMT is also studied by solving the
corresponding eigenvalue problem. 
 As an application
of the developed formulas, in Sec. IV
we consider a stationary shell that generates the Lewis solutions, 
which represent the most general cylindrical vacuum solutions of
the Einstein field equations with 
rotation \cite{Lewis1932}, while in Sec. V our main conclusions are given.

\section{Cylindrical Spacetimes with Rotation}

To begin with,  let us consider the cylindrical spacetimes with rotation 
described by the metric \cite{MQ1990},
\begin{equation}
\lb{eq1}
ds^{2}=e^{2(\chi-\psi)}\left(dt^{2}-dr^{2}\right)-W^{2} 
e^{-2\psi}\left(\omega dt+d\varphi\right)^{2}-e^{2\psi}dz^{2},
\end{equation}
where $\psi$, $\chi$, $W$ and $\omega$ are functions  $t$ and $r$,
and $\{x^{\mu}\}\equiv \{t, r, z, \varphi\}, \; (\mu = 0, 1, 2, 3)$,
are the usual cylindrical coordinates. In general the spacetimes have
two Killing vectors, one is associated with the invariant translations
along the symmetry axis,
$\xi_{(z)} = \partial {z}$, and the other is associated with
the invariant rotations  about the axis, $\xi_{(\varphi)} = 
\partial {\varphi}$. Clearly, for the metric given above,
the two Killing vectors are  orthogonal. 
When $\omega = 0$,  the  metric represents spacetimes
without rotation, in which the polarization of gravitational waves  
has only one degree of freedom and the  direction of polarization is fixed
\cite{Thorne1965,MTW1973}. For the spacetimes to be 
cylindrical, several criteria have to be satisfied 
\cite{PSW1996}. When the symmetry axis is regular, those
conditions are easily imposed. However, when it is singular,
it is still not clear  which conditions should be imposed
\cite{MS1998}.

Corresponding to the metric (\ref{eq1}) the 
Christoffel symbols and Einstein tensor are given,
respectively, by Eqs.(\ref{b1}) and (\ref{b2}). 
To further study the
spacetimes, let us consider the Ricci and Weyl scalars,
as these quantities have their explicit physical interpretations
\cite{NP1962}, and they are particularly
useful in the study of gravitational interaction among
mater fields and gravitational waves \cite{Szekeres1965,Wang1991}. 
Choosing the  null tetrad \footnote{It should be noted that
the choice of the null tetrad used here is different from the one  
used in \cite{LW1994} for the case $\omega = 0$, 
and can be obtained one from the other
by exchanging the roles of the two null vectors, $l_{\mu}$ and 
$n_{\mu}$.}
\begin{eqnarray}
\lb{eq2}
l^{\mu} &=& \frac{e^{\psi - \chi}}{\sqrt{2} A}
\left\{1, \;1, \;0, \; - \omega\right\},\;\;
n^{\mu} = \frac{A e^{\psi - \chi}}{\sqrt{2}}
\left\{1, \; - 1, \;0, \; - \omega\right\},\nb\\
m^{\mu} &=& \frac{1}{\sqrt{2}}
\left\{0, \; 0, \; e^{-\psi}, \; i \frac{e^{\psi}}{W}\right\},
\;\;
\bar{m}^{\mu} = \frac{1}{\sqrt{2}}
\left\{0, \; 0, \; e^{-\psi}, \; - i \frac{e^{\psi}}{W}\right\},
\end{eqnarray}
where a bar denotes the complex conjugate, and $A$ is an arbitrary 
function of $t$ and $r$,
we find that the spin coefficients are given by Eq.(\ref{a1}).
Since $\kappa = \nu = 0$, the two null vectors $ l^{\mu}$ and $n^{\mu}$
 define two null geodesic congruences
  \cite{Frolov1979}. The one defined by
 $l_{\mu}$ is outgoing, while 
the one defined by $n_{\mu}$ is ingoing, similar to the no-rotating 
case \cite{LW1994}. In fact, it is
easy to show that  
\bq
\lb{eq3}
l_{\mu;\nu}l^{\nu} = 2 \epsilon l_{\mu},\;\;\;\;
n_{\mu;\nu}n^{\nu} = 2\gamma n_{\mu},
\eq
where $\epsilon$ and $\gamma$ are given by Eq.(\ref{a1}). Choosing
the function $A$ as $A = e^{\chi-\psi}$, from Eq.(\ref{a1})
we find that $\epsilon = 0$, and then Eq.(\ref{eq3}) shows
that the null geodesic congruence defined by $l_{\mu}$ is affinely 
parameterized. Hence, the corresponding 
expansion, rotation and shear of the outgoing
null geodesic congruence are given, respectively, by
\bqn
\lb{eq4}
\theta_{l} &\equiv& \frac{1}{2} l^{\mu}_{\;\; ;\mu} = - Re(\rho) =  
\frac{e^{2(\psi - \chi)}}{2\sqrt{2}}\frac{W_{,t} + W_{,r}}{W},\nb\\
\omega^{2}_{l} &\equiv& \frac{1}{2}l_{[\mu;\nu]}l^{\mu;\nu} 
= \left[Im(\rho)\right]^{2} = 0,\nb\\
\sigma_{l} &\equiv& \left(\frac{1}{2}l_{(\mu;\nu)}l^{\mu;\nu} 
- \theta^{2}_{l}\right)^{1/2} 
= \frac{e^{2(\psi - \chi)}}{2\sqrt{2}}
\left[2(\psi_{,t} + \psi_{,r}) - \frac{W_{,t} + W_{,r}}{W}\right],
\eqn
where 
\bq
\lb{eq4a}
A_{(\mu\nu)} \equiv \frac{1}{2}\left(A_{\mu\nu} + A_{\nu\mu}\right),\;\;
A_{[\mu\nu]} \equiv \frac{1}{2}\left(A_{\mu\nu} - A_{\nu\mu}\right),
\eq
and $W_{,t} \equiv \partial W/ \partial t,\; W_{,r} \equiv \partial
W/\partial r$, etc.
Similarly, if we choose the arbitrary function $A$ 
as $A = e^{\psi -\chi}$, Eq.(\ref{a1})
yields $\gamma = 0$, then Eq.(\ref{eq3}) shows
that the null geodesic congruence defined by $n_{\mu}$ now is affinely 
parameterized, and the corresponding expansion, rotation and shear of 
the ingoing null geodesic congruence are given, respectively, by
\bqn
\lb{eq5}
\theta_{n} &\equiv& \frac{1}{2} n^{\mu}_{\;\; ;\mu} = Re(\mu) =  
\frac{e^{2(\psi - \chi)}}{2\sqrt{2}}\frac{W_{,t} - W_{,r}}{W},\nb\\
\omega^{2}_{n} &\equiv& \frac{1}{2}n_{[\mu;\nu]}n^{\mu;\nu} 
= \left[Im(\mu)\right]^{2} = 0,\nb\\
\sigma_{n} &\equiv& \left(\frac{1}{2}n_{(\mu;\nu)}n^{\mu;\nu} 
- \theta^{2}_{n}\right)^{1/2}
= \frac{e^{2(\psi - \chi)}}{2\sqrt{2}}
\left[2(\psi_{,t} - \psi_{,r}) - \frac{W_{,t} - W_{,r}}{W}\right].
\eqn

Once the spin coefficients are given, we can calculate the 
corresponding Ricci and Weyl scalars, which    
are given, respectively, 
by Eqs.(\ref{a2}) and (\ref{a3}). From those expressions 
we can see that all the Weyl scalars 
are non-zero, and each of them has the following physical 
interpretation \cite{Szekeres1965,Wang1991}: 
The $\Psi_{0}$ and $\Psi_{1}$
terms represent, respectively, the 
transverse and longitudinal gravitational
wave components along the null geodesic
 congruence defined by $n^{\mu}$,
and the $\Psi_{4}$ and $\Psi_{3}$
terms represent, respectively, the transverse 
and longitudinal gravitational
wave components along the null geodesic 
congruence defined by $l^{\mu}$,
while the $\Psi_{2}$
term represents the ``Coulomb" component.

The physical meaning of the Weyl and Ricci
scalars can be further studied from
 geodesic deviations. Because of the symmetry, it is sufficient only to
consider the null geodesics defined by $l^{\mu}$,
which are affinely parametrized when $A = e^{\chi - \psi}$, as shown above. 
Let $\eta^{\mu}$ be the 
deviation vector between two neighbor geodesics, and $\eta^{\mu}l_{\mu}
 = 0$. Then, using Eqs.(\ref{a2}) -(\ref{a7}), we find that  
the geodesic deviation can be
written in the form
\bqn
\lb{eq6}
\frac{D^{2}\eta^{\mu}}{D\lambda^{2}}
&=& - R^{\mu}_{\nu\alpha\beta}l^{\nu}l^{\beta}\eta^{\alpha}\nb\\
&=& \left\{ \Phi_{00}e^{\mu\nu}_{0} + \Psi_{0}e^{\mu\nu}_{+} 
+ i \left(\Psi_{1}+ \Phi_{01}\right)
e_{03}^{\mu\nu} 
+ i \left(\Psi_{1}+ \Phi_{01}\right) e_{13}^{\mu\nu}
\right\}\eta_{\nu},
\eqn
where $\Psi_{0}, \; \Psi_{1},\; \Phi_{00}$, and
$\Phi_{01}$ are given by Eqs.(\ref{a3}) and (\ref{a4})
with $A = e^{\chi - \psi}$, and 
\bqn
\lb{eq7}
e^{\mu\nu}_{0} &\equiv& e^{\mu}_{2}e^{\nu}_{2} 
+ e^{\mu}_{3}e^{\nu}_{3},\;\;
e^{\mu\nu}_{+} \equiv e^{\mu}_{2}e^{\nu}_{2} 
- e^{\mu}_{3}e^{\nu}_{3},\nb\\
e^{\mu\nu}_{03} &\equiv& e^{\mu}_{0}e^{\nu}_{3} 
+ e^{\mu}_{3}e^{\nu}_{0},\;\;
e^{\mu\nu}_{13} \equiv e^{\mu}_{1}e^{\nu}_{3} 
+ e^{\mu}_{3}e^{\nu}_{1},
\eqn
with
\bqn
\lb{eq8}
e^{\mu}_{0} &\equiv& \frac{l^{\mu} + n^{\mu}}{\sqrt{2}},\;\;
e^{\mu}_{1} \equiv \frac{l^{\mu} - n^{\mu}}{\sqrt{2}},\nb\\
e^{\mu}_{2} &\equiv& \frac{m^{\mu} + \bar{m}^{\mu}}{\sqrt{2}},\;\;
e^{\mu}_{3} \equiv - \frac{i\left({m}^{\mu} - 
\bar{m}^{\mu}\right)}{\sqrt{2}}.
\eqn
Eqs.(\ref{eq6}) has the following physical interpretation 
\cite{Szekeres1965,Wang1991}. 
Let $S_{O}$ and $S_{P}$ be infinitesimal 2-elements spanned by $e^{\mu}_{2}$
and $e^{\mu}_{3}$ and orthogonal to a null geodesic $C$ defined by $l^{\mu}$,
passing $S_{O}$ and $S_{P}$ at the points $O$ and $P$, respectively. Let
$S$ be an infinitesimal circle with center $O$, lying in $S_{O}$ as
illustrated by Fig.1. Suppose 
that a light beam meets $S_{O}$ in the circle $S$, then each
of them has the following effect on  
the image of the circle $S$ on $S_{P}$. The first term 
$\Phi_{00}$ in Eq.(\ref{eq6}) will always make the circle contracted, 
as for any 
physically realistic matter field we have $\Phi_{00} \ge 0$ [cf. 
Fig.2(a)]. The second term $\Psi_{0}$ will make the circle 
elliptic with the main major axis along $e^{\mu}_{2}$, as 
shown by Fig.2(b). 
To see the physical interpretation of the last two terms, let us consider
a tube along the null geodesic $C$.  Consider a sphere consisting photons, 
which will cut $S_{O}$ in the circle $S$ with the point $O$ as its center,
as shown in Fig.3. Then, the last term in Eq.(\ref{eq6})
will make the image of the sphere at the point $P$ as a 
spheroid with the main major axis along a line at $45^{0}$ 
with respect to $e^{\mu}_{1}$ in the plane spanned by $e^{\mu}_{1}$ 
and $e^{\mu}_{3}$, while the rays are left undeflected  in the 
$e^{\mu}_{2}$-direction. This can be seen clearly by performing 
a rotation in the plane spanned by $e^{\mu}_{1}$ 
and $e^{\mu}_{3}$,
\bq
\lb{eq9}
e^{\mu}_{1} = \cos\alpha {e'}_{1}^{\mu} - \sin\alpha {e'}_{3}^{\mu},\;\;
e^{\mu}_{3} = \sin\alpha {e'}_{1}^{\mu} + \cos\alpha {e'}_{3}^{\mu},
\eq
which leads to
\bq
\lb{eq10}
e^{\mu\nu}_{13} = \sin(2\alpha)({e'}_{1}^{\mu}{e'}_{1}^{\nu} 
- {e'}_{3}^{\mu}{e'}_{3}^{\nu})
 + \cos(2\alpha)({e'}_{1}^{\mu}{e'}_{3}^{\nu} 
+ {e'}_{3}^{\mu}{e'}_{1}^{\nu}).
\eq
Thus, choosing $\alpha = \pi/4$, we have
\bq
\lb{eq11}
e^{\mu\nu}_{13} = {e'}_{1}^{\mu}{e'}_{1}^{\nu} 
- {e'}_{3}^{\mu}{e'}_{3}^{\nu},\; (\alpha = \pi/4).
\eq
Combining Eq.(\ref{eq6}) and the above equation, we can see that
the last term will make a circle in the ${e'}_{1}{e'}_{3}$-plane into
an ellipse with its main major axis along the ${e'}_{1}$-axis, which
is at $45^{0}$ with respect to the $e_{1}$-axis. 
It should be noted that in the case of timelike geodesics
the $\Psi_{1}$ term  deflects the sphere into an ellipsoid
\cite{Szekeres1965}. 
Moreover, the third term in Eq.(\ref{eq6})
is absent in the timelike geodesic case. The effect of this term 
will make a clock ``flying" with the photons slow down, in addition to
the effect of deflecting the photons in the $e^{\mu}_{3}$-direction.
It is interesting to note that there is a fundamental difference between
the time delay caused by this term and the one caused by a Lorentz 
boost. The latter, in particular, has no contribution to geodesic
deviations, timelike or null.
From the above analysis we can see that for a pure Petrov type $N$
gravitational wave
propagating along the null geodesic congruence, in which only
the component $\Psi_{0}$ is different from zero, the
gravitational wave has only one polarization state,
similar to the case without rotation 
\cite{Thorne1965,Szekeres1965,Wang1991}. The difference between
these two cases is that in the case
without rotation, the polarization angle remains the same even in 
different points along the wave path, 
while when $\omega\not= 0$, in general
this is no longer true. In fact, it is easy to 
show that 
\bq
\lb{eq12}
e^{\mu}_{2;\nu}l^{\nu}  = 0,\;\;
e^{\mu}_{3;\nu}l^{\nu} 
= - W e^{\psi - \chi}\frac{\omega,_{r}}{2\sqrt{2}}
 \left(e^{\mu}_{0} + e^{\mu}_{1}\right).
\eq
Thus, although $e^{\mu}_{2}$ is parallel-transported along the null
geodesic congruence, $e^{\mu}_{3}$ in general is not, and is rotating
with respect to a parallel-transported basis. 
Since the polarization angle of
the $\Psi_{0}$ wave remains the same with respect to $e^{\mu}_{3}$,
the polarization direction
is also rotating with respect to the parallel-transported basis.

 %%%%%%%%%%%%%%%%%%%%%%%%%%%%%%%%%%%%%%%%%%%%%%%%%%%%%%%%%%%%%%%
\section{Rotating Cylindrical Shells}
%%%%%%%%%%%%%%%%%%%%%%%%%%%%%%%%%%%%%%%%%%%%%%%%%%%%%%%%%%%%%%%%%%

In the last section, the main properties of the spacetimes with rotation
have been studied. Since the Einstein field equations  
 are all involved with derivatives of the metric coefficients
up to the second order, and the Bianchi identities, which are
usually considered as representing  the 
interaction among gravitational fields and matter fields, up 
to the third
order, it is generally assumed that the metric coefficients
are at least $C^{3}$ \cite{HE1973}, that is, 
the derivatives of the metric
coefficients exist and continuous at least up to their third order.
 However, for a long time it has been 
realized that this condition is too strict and rules out many physically
interesting cases, such as, shells, star, and so on.

In this section, we shall generalize the formulas given in the last
section to the case where the metric coefficients are $C^{3}$ only
in certain regions, while across the boundaries that separate these
regions they are $C^{0}$, that is, the metric coefficients are only 
continuous across the boundaries. These boundaries can be classified into two 
different kinds, one is boundary surfaces, like a star,
the other is surface layers, like a matter shell
\cite{Israel1966}. In the former case, the extrinsic curvatures of
the boundaries in their two faces are equal, while in the latter case
they are not, and as a result matter shells in general
appear on these boundaries. In this paper, we shall treat 
the two cases together and consider the former is a particular 
case of the latter.  

In a given spacetime, there may exist many disconnected boundaries. 
However, in the following we shall consider the case where there exists
only one boundary, as its generalization to the cases of many 
boundaries is straightforwards. Assume that the whole spacetime is divided
into two regions $V^{\pm}$ by a hypersurface $\Sigma$, where
\bq
\lb{eq13}
V^{+} = \{x^{\mu}: \phi > 0\}, \;\;\;
V^{-} = \{x^{\mu}: \phi < 0\},\;\;\;
\Sigma = \{x^{\mu}: \phi = 0\}, 
\eq
with
\bq
\lb{eq14}
\phi = r - R(t),
\eq
where $R(t)$ is an arbitrary function that describes the history of the
boundary. Then, for any of the metric coefficients, which is $C^{0}$ across 
the boundary and $C^{3}$ in the regions $V^{\pm}$ can be written 
in the form \cite{Wang1992}
\bq
\lb{eq15}
f(t,r) = f^{+}(t,r)H(\phi) + f^{-}(t,r)[1 - H(\phi)],
\eq
where $f = \{\psi, \;\chi, \; W, \; \omega\}$,
$H(\phi)$ is the Heavside function defined as \footnote{It should
be noted that the exact value of $H(\phi)$ at the point $\phi = 0$
is not uniquely defined and can be given any value.}
\bq
\lb{eq16}
H(\phi)= \cases{
1, & $\phi \ge 0$,\cr
0, & $\phi > 0$, \cr}
\eq
and $f^{+}\; (f^{-})$ is the function defined in the region 
$V^{+}\; (V^{-})$, with the $C^{0}$ condition
\bq
\lb{eq17}
\lim_{r \rightarrow R^{+}}{ f^{+}(t,r)} =
\lim_{r \rightarrow R^{-}}{ f^{-}(t,r)}.
\eq
Using the distribution theory, it can be shown that 
\bqn
\lb{eq18}
f_{,t}(t,r) &=& f_{,t}^{+}(t,r)H(\phi) 
+ f_{,t}^{-}(t,r)[1 - H(\phi)],\nb\\ 
f_{,r}(t,r) &=& f_{,r}^{+}(t,r)H(\phi) 
+ f_{,r}^{-}(t,r)[1 - H(\phi)],\nb\\
f_{,tt}(t,r) &=& f_{,tt}^{+}(t,r)H(\phi) 
+ f_{,tt}^{-}(t,r)[1 - H(\phi)]
+ \dot{R}(t)^{2}[f_{,r}]^{-}\delta(\phi),\nb\\
f_{,tr}(t,r) &=& f_{,tr}^{+}(t,r)H(\phi) 
+ f_{,tr}^{-}(t,r)[1 - H(\phi)]
- \dot{R}(t)[f_{,r}]^{-}\delta(\phi),\nb\\
f_{,rr}(t,r) &=& f_{,rr}^{+}(t,r)H(\phi) 
+ f_{,rr}^{-}(t,r)[1 - H(\phi)]
+ [f_{,r}]^{-}\delta(\phi),
\eqn
where an over-dot denotes the ordinary differentiation with 
respect to $t$, $\delta(\phi)$ the Dirac delta function, 
and
\bq
\lb{eq19}
[f_{,r}]^{-} \equiv \lim_{r \rightarrow R(t)^{+}}\left(
        \frac{\partial f^{+}(t, r)}{\partial r}\right) - 
        \lim_{r \rightarrow R(t)^{-}}\left(
        \frac{\partial f^{-}(t, r)}{\partial r}\right).
\eq
It should be noted that in deriving Eq.(\ref{eq18}) we have used the
relation
\bq
\lb{eq20}
[f_{,t}]^{-}  = - \dot{R}(t)[f_{,r}]^{-}.
\eq

Substituting Eq.(\ref{eq18}) into Eq.(\ref{b2}), we find that the
Einstein tensor in general can be written in the form
\bq
\lb{eq21}
G_{\mu\nu} = G^{+}_{\mu\nu}H(\phi) 
            + G^{-}_{\mu\nu}[1 - H(\phi)] 
            + \gamma_{\mu\nu}\delta(\phi),
\eq
where $G^{+}_{\mu\nu}\; (G^{-}_{\mu\nu})$ is the Einstein tensor
calculated in the region $V^{+}\; (V^{-})$, and $\gamma_{\mu\nu}$ is the
Einstein tensor calculated on the hypersurface $r = R(t)$. The non-vanishing
components of $\gamma_{\mu\nu}$ in the present case are given by
\bqn
\lb{eq22}
\gamma_{00}  &=& - \frac{[W_{r}]^{-}}{W_{0}} 
              + \omega_{0}^{2}W_{0}^{2}e^{-2\chi_{0}}\left\{
              (1 - \dot{R}^{2})[\chi_{,r}]^{-} -
              \frac{[\omega_{,r}]^{-}}{\omega_{0}}\right\},\nb\\
\gamma_{01}  &=& \dot{R}\left\{ \frac{[W_{r}]^{-}}{W_{0}} 
              + \frac{1}{2}\omega_{0}W_{0}^{2}e^{-2\chi_{0}} 
              [\omega_{,r}]^{-}\right\},\nb\\
\gamma_{03}  &=& \frac{1}{2}\omega_{0}W_{0}^{2}e^{-2\chi_{0}}
                \left\{ 2(1 - \dot{R}^{2})[\chi_{,r}]^{-} -
              \frac{[\omega_{,r}]^{-}}{\omega_{0}}\right\},\nb\\
\gamma_{11}  &=& - \dot{R}^{2} \frac{[W_{,r}]^{-}}{W_{0}},\nb\\
\gamma_{13}  &=& \frac{1}{2}\dot{R}W_{0}^{2}e^{-2\chi_{0}}
                 [\omega_{,r}]^{-},\nb\\              
\gamma_{22}  &=& (1 - \dot{R}^{2})e^{2(2\psi_{0}-\chi_{0})}
                 \left\{[\chi_{,r}]^{-} - 2[\psi_{,r}]^{-} +
              \frac{[W_{,r}]^{-}}{W_{0}}\right\},\nb\\                    
\gamma_{33}  &=& (1 - \dot{R}^{2})W_{0}^{2}e^{-2\chi_{0}}
                 [\chi_{,r}]^{-}, \;(r = R(t)),
\eqn
where the quantities with the subscript ``$0$" denote the ones calculated
on the hypersurface $r = R(t)$, for example, $\omega_{0} \equiv 
\omega(t, \; R(t))$, and so on. Writing the energy-momentum tensor (EMT)
in a form of Eq.(\ref{eq21}), we find that the Einstein field equations
$G_{\mu\nu} = kT_{\mu\nu}$, where $k$ is the Einstein constant, can
be written as 
\bqn
\lb{eq23}
G^{+}_{\mu\nu} &=& k T^{+}_{\mu\nu},\; (r > R(t)), \\
G^{-}_{\mu\nu} &=& k T^{-}_{\mu\nu},\; (r < R(t)), \\
\gamma_{\mu\nu} &=& k \tau_{\mu\nu},\; (r = R(t)),
\eqn
where $\tau_{\mu\nu}$ can be interpreted as representing the surface
energy-momentum  tensor, which in the present case 
takes the form
\bq
\lb{eq24}
\tau_{\mu\nu} = \eta u_{\mu}u_{\nu} + p_{z}z_{\mu}z_{\nu}
 + p_{\varphi}\varphi_{\mu}\varphi_{\nu} + q(u_{\mu}\varphi_{\nu}
 + u_{\nu}\varphi_{\mu}),
\eq
where
\bqn
\lb{eq25}
\eta &\equiv& - (1 - \dot{R}^{2})e^{2(\psi_{0}-\chi_{0})}
               \frac{[W_{,r}]^{-}}{W_{0}},\nb\\ 
q  &\equiv& - \frac{1}{2}(1 - \dot{R}^{2})^{1/2}
                W_{0}e^{2\psi_{0}-3\chi_{0}}
                 [\omega_{,r}]^{-}, \nb\\      
p_{z} &\equiv&  (1 - \dot{R}^{2})e^{2(\psi_{0}-\chi_{0})}
               \left\{[\chi_{,r}]^{-} - 2[\psi_{,r}]^{-} +
              \frac{[W_{,r}]^{-}}{W_{0}}\right\},\nb\\  
p_{\varphi} &\equiv& (1 - \dot{R}^{2})e^{2(\psi_{0}-\chi_{0})}
                [\chi_{,r}]^{-}, \;(r = R(t)),    
\eqn
and
\bqn
\lb{eq26}
u_{\mu} &=& (1 - \dot{R}^{2})^{-1/2}e^{\chi_{0} - \psi_{0}}\left\{
        \delta^{t}_{\mu} - \dot{R}\delta^{r}_{\mu}\right\},\nb\\
\eta_{\mu} &=& (1 - \dot{R}^{2})^{-1/2}e^{\chi_{0} - \psi_{0}}\left\{
        \delta^{r}_{\mu} - \dot{R}\delta^{t}_{\mu}\right\},\nb\\
z_{\mu} & = & e^{\psi_{0}}\delta^{z}_{\mu},\;\;
\varphi_{\mu}  =  W_{0} e^{-\psi_{0}}\left\{\omega_{0}\delta^{t}_{\mu}
+ \delta^{\varphi}_{\mu}\right\},
\eqn
with the properties
\bqn
\lb{eq27}
u_{\lambda}u^{\lambda} &=& - z_{\lambda}z^{\lambda}
= - \varphi_{\lambda}\varphi^{\lambda} = 1,\nb\\
u_{\lambda}z^{\lambda} &=& - u_{\lambda}\varphi^{\lambda}
= - z_{\lambda}\varphi^{\lambda} = 0.
\eqn

In order to have the physical interpretation for each term appearing in
Eq.(\ref{eq24}), we need to cast the surface EMT in its canonical form
\cite{HE1973,LW94}, that is, we need to solve the eigenvalue problem,
\bq
\lb{eq27a}
\tau^{\mu}_{\nu} \xi^{\nu} = \lambda \xi^{\mu}.
\eq
This system of equations will possess nontrivial solutions only when the
determinant $det|\tau^{\mu}_{\nu} - \lambda \delta^{\mu}_{\nu}| = 0$, which
in the present case can be written as
\bq
\lb{eq27b}
\lambda (p_{z} - \lambda)\left[\lambda ^{2} - (\eta -
p_{\varphi})\lambda + q^{2} - \eta p_{\varphi}\right] = 0.
\eq 
Clearly, the above equation has four roots,
$
\lambda = 0,\;  p_{z},\; \lambda _{\pm},
$
where
\bq
\lb{eq27d}
\lambda _{\pm} = \frac{1}{2} \left[(\eta - p_{\varphi}) \pm
D^{1/2}\right],\;\;\;
 D \equiv (\eta + p_{\varphi})^{2} - 4 q^{2}.
\eq
It can be shown that the eigenvalue $\lambda = 0$ corresponds to the
eigenvector $\xi_{1} ^{\mu} = \eta^{\mu}$, where $\eta^{\mu}$ is the normal
vector to the hypersurface $r = R(t)$, and given by Eq.(\ref{eq26}). The
eigenvalue $\lambda = p_{z}$ corresponds to the eigenvector $\xi_{2}^{\mu} =
z^{\mu}$, which represents the pressure of the shell in the $z$-direction.

On the other hand, substituting Eq.(\ref{eq27d}) into Eq.({\ref{eq27a}), we
find that the corresponding eigenvectors are given, respectively, by
\bq
\lb{eq27f}
\xi^{\mu}_{\pm} = (\lambda_{\pm} + p_{\varphi})u^{\mu} + q \varphi^{\mu}.
\eq
To further study the physical meaning of $\lambda _{\pm}$, it is found
convenient to distinguish the three cases: (a) $D > 0$; (b) $D = 0$; and
(c) $D < 0$.

{\bf Case (a)}: In this case, the two roots $\lambda _{\pm}$ and
the two corresponding eigenvectors $\xi_{\pm}^{\mu}$ are all real and  satisfy
the relations, 
\bqn
\lb{eq27g}
(\lambda_{+} + p_{\varphi})(\lambda_{-} + p_{\varphi}) &=& q^{2},\nb\\
\frac{\xi^{\mu}_{\pm} \xi^{\nu}_{\pm} g_{\mu\nu} }
{D^{1/2}(\lambda_{\pm} + p_{\varphi})} &=& \pm 1,\nb\\
\xi^{\mu}_{+} \xi^{\nu}_{-} g_{\mu\nu} &=& 0.
\eqn
From these expressions we can see that when $\lambda_{+} + p_{\varphi} > 0$,
the eigenvector $\xi_{+}^{\mu}$ is timelike, while   $\xi_{-}^{\mu}$
is spacelike. Setting
\bqn
\lb{eq27h}
E_{(0)}^{\mu} &\equiv& \frac{\xi^{\mu}_{+}}{\left[D^{1/2}(\lambda_{+} +
p_{\varphi})\right]^{1/2}},\nb\\
E_{(3)}^{\mu} &\equiv& \frac{\xi^{\mu}_{-}}{\left[D^{1/2}(\lambda_{-} +
p_{\varphi})\right]^{1/2}},\;(\lambda_{+} + p_{\varphi} > 0),
\eqn
we find that $E_{(a)} ^{\mu},\; (a = 0,1,2,3)$ form an orthogonal base,
i.e., $E^{\lambda}_{(a)}E_{(b)\lambda} = \eta_{ab}$, where
\bq
\lb{eq27ha}
E_{(1)}^{\mu} = \eta^{\mu},\;\;\; E_{(2)}^{\mu} = z^{\mu}.
\eq
Then, in terms of these unit vectors, the surface EMT given by
Eq.(\ref{eq24}) takes the form
\bq
\lb{eq27i}
\tau^{\mu\nu} = \Sigma E_{(0)}^{\mu}E_{(0)}^{\nu} 
+ p_{z}E_{(3)}^{\mu}E_{(3)}^{\nu}
+ p_{(3)}E_{(3)}^{\mu}E_{(3)}^{\nu},  
\eq
where
\bqn
\lb{eq27j}
\Sigma &\equiv& \frac{D\left(\lambda_{+} + p_{\varphi}\right)}{2q^{2}}
\left\{ D^{1/2}p_{\varphi} - \left[p_{\varphi}(\eta + p_{\varphi}) -
2q^{2}\right]\right\},\nb\\
p_{(3)} &\equiv&\frac{D\left(\lambda_{-} + p_{\varphi}\right)}{2q^{2}}
\left\{ D^{1/2}p_{\varphi} + \left[p_{\varphi}(\eta + p_{\varphi}) -
2q^{2}\right]\right\}, \; (\lambda_{+} + p_{\varphi} > 0).
\eqn
Hence, in terms of its tetrad components, $\tau^{\mu\nu}$ can be cast in the
form, 
\bq
\lb{eq27k}
\left[\tau_{(a)(b)}\right] = 
\left[\matrix{ 
\Sigma & 0 & 0 & 0 \cr
0 & 0 & 0 & 0 \cr
0 & 0 & p_{z} & 0 \cr
0 & 0 & 0 & p_{(3)} \cr }\right].
\eq
This corresponds to the Type I fluid defined in \cite{HE1973}.
Thus, in this case the surface EMT represents a fluid with its surface energy
density given by $\Sigma$, measured by observers whose four-velocity
are given by $E_{(0)}^{\mu}$, and the principal pressures in the  
directions $E_{(2)}^{\mu}$ and $E_{(3)}^{\mu}$, given respectively by $p_{z}$
and $p_{(3)}$.

When $\lambda_{+} + p_{\varphi} < 0$, the eigenvector $\xi_{+}^{\mu}$ is
spacelike, while   $\xi_{-}^{\mu}$ is timelike. Now if we define the two unit
vectors $E_{(0)}$ and $E_{(3)}$ as
\bqn
\lb{eq27l}
E_{(0)}^{\mu} &\equiv& \frac{\xi^{\mu}_{-}}{\left[D^{1/2}\left|\lambda_{-} +
p_{\varphi}\right|\right]^{1/2}},\nb\\
E_{(3)}^{\mu} &\equiv& \frac{\xi^{\mu}_{+}}{\left[D^{1/2}\left|\lambda_{+} +
p_{\varphi}\right|\right]^{1/2}},\; (\lambda_{+} + p_{\varphi} < 0),
\eqn
we find that the surface EMT also takes the form of Eq.(\ref{eq27k}), but now
with
\bqn
\lb{eq27m}
\Sigma &\equiv&\frac{D\left|\lambda_{-} + p_{\varphi}\right|}{2q^{2}}
\left\{ D^{1/2}p_{\varphi} + \left[p_{\varphi}(\eta + p_{\varphi}) -
2q^{2}\right]\right\},\nb\\
p_{(3)} &\equiv&\frac{D\left|\lambda_{+} + p_{\varphi}\right|}{2q^{2}}
\left\{ D^{1/2}p_{\varphi} - \left[p_{\varphi}(\eta + p_{\varphi}) -
2q^{2}\right]\right\}, \; (\lambda_{+} + p_{\varphi} < 0).
\eqn

{\bf Case (b)}: In this case we have
\bq
\lb{eq27n}
q = \pm\frac{1}{2}(\eta - p_{\varphi}).
\eq
Then, the two roots $\lambda_{\pm}$ degenerate
into one. It can be shown that this multiple root corresponds to two
independent eigenvectors,
\bq
\lb{eq27o}
\xi^{\mu}_{\pm} = \frac{u^{\mu} \pm \varphi^{\mu}}{\sqrt{2}},
\eq
which are all null. From these two null vectors we
can construct two unit vectors, one is timelike and the other is spacelike.
But, these two unit vectors are exactly $u^{\mu}$ and  $\varphi^{\mu}$. Then,
in the base $E_{(a)}^{\mu} = \left\{u^{\mu},\; \eta^{\mu},\; z^{\mu},\;
\varphi^{\mu}\right\}$, the surface EMT will  take the form 
\bq
\lb{eq27p}
\left[\tau_{(a)(b)}\right] = 
\left[\matrix{ 
\Omega + \Sigma  & 0 & 0 & \pm \Omega\cr
0& 0 & 0 & 0\cr
0& 0 & p_{z}& 0 \cr
\pm \Omega& 0& 0& \Omega - \Sigma \cr}
\right],
\eq
 with  
\bq
\lb{eq27q}
\Sigma \equiv \frac{1}{2}(\eta - p_{\varphi}),\;\;\;
\Omega \equiv \frac{1}{2}(\eta + p_{\varphi}).
\eq
This corresponds to the Type II fluid defined in \cite{HE1973}.

{\bf Case (c)}: In this case the two roots $\lambda_{\pm}$ are complex, and
satisfy  the relations $\lambda _{-} = \bar{\lambda}_{+}$. The two
corresponding eigenvectors, given by Eq.(\ref{eq27f}),  now are also complex.
This means that in the present case the surface EMT   
cannot be diagonalized (by real similarity transformations), and is already in
its canonical form. Thus, in the base consisting of the four orthogonal
vectors, $E^{\mu}_{(a)} = \{u^{\mu}, \eta^{\mu}, z^{\mu}, \varphi^{\mu}\}$, it
takes the form, \bq
\lb{eq27r}
\left[\tau_{(a)(b)}\right] = 
\left[
\matrix{ 
\eta  & 0 & 0 & q\cr
0& 0 & 0 & 0\cr
0& 0 & p_{z}& 0 \cr
q & 0& 0& p_{\varphi} \cr}
\right],
\eq
from which we can see that now  $\eta$ denotes the surface energy density of
the shell, measured by observers whose four-velocity are given by
$u^{\mu}$, $p_{z}$ and $p_{\varphi}$ the principal pressures in the directions,
$z^{\mu}$ and $\varphi^{\mu}$, respectively, and $q$ the heat flow in the  
$\varphi^{\mu}$-direction. 
  
It should be noted that all the above physical interpretations are valid, 
provided that the surface EMT satisfies some energy conditions \cite{HE1973}.

\section{Stationarily Rotating Shells}

As an application of the formulas developed in the last section, 
in this section we shall consider  stationarily rotating
cylindrical shells. It should be noted that such shells were studied
previously by various authors \cite{Fr1972}. However, an important difference
in the present case is that now we allow the spacetime inside the shell be
rotating,    
\bq
\lb{4.1}
ds^{2}_{-} = dt^{2} - dr^{2} - r^{2}(d\varphi + \Omega dt)^{2}
- dz^{2},
\eq
where $\Omega$ is a constant that represents the angular velocity of
the uniformly rotating coordinate system \cite{LL1975}. 
Note that in the above metric the axis $r = 0$ is well-defined 
and free of any kind of singularities and sources. 
When $r > \Omega^{-1}$,  the metric
coefficient $g^{-}_{00}$ becomes negative and
the Killing vector $\xi_{(0)}^{\mu} = 
\delta^{\mu}_{0}$ becomes space-like. In the following we shall
assume that  the above metric is valid only for $r < \Omega^{-1}$.

If we further
assume that outside of the shell, the spacetime is vacuum, then  
it should be described by the most
general vacuum Lewis solutions of the Einstein field
equations, given by \cite{Kramer1980}
\bq
\lb{4.2}
ds^{2}_{+} = f d\bar{t}^{2} + 2k d\bar{t}d\bar{\varphi} - h(d\bar{r}^{2}
+ d\bar{z}^{2}) - l d\bar{\varphi}^{2},
\eq
where
\bqn
\lb{4.3}
f&=& a\bar{r}^{1 - n} - \frac{c^{2}}{an^{2}}\bar{r}^{1 + n},\;\;
k = Af,\;\;\;
h = \bar{r}^{(n^{2} - 1)/2},\nb\\
l &=& \frac{\bar{r}^{2}}{f} - A^{2}f,\;\;\;
A = \frac{c\bar{r}^{1+n}}{anf} + b,
\eqn
with $a, \; b,\; c$ and $n$ being constants that can be real
or complex. When they are complex, certain relations have to be
satisfied among them, in order to have the metric coefficients 
be real \cite{Lewis1932,Fatima1995}. For the present purpose, 
we shall consider only the case where they are all real.
 
It should be noted that the above expressions are
valid only for $n \not= 0$. However, the solutions for $n = 0$ 
can be obtained from them by first letting $c = n\bar{c}$ 
and then taking the limit $n \rightarrow 0$. Since this process is 
straightforwards, in the following we shall consider only the solutions
given by Eqs.(\ref{4.2}) and (\ref{4.3}), and consider the ones
for $n = 0$ as their particular case. The physical meaning of 
the parameters $a, \; b,\; c$ and $n$ were first studied by Lewis 
 \cite{Lewis1932}, and more recently by da Silva {\em et al}. 
\cite{Fatima1995}. 
When all the parameters are real, it can be shown that  
the spacetime is
asymptotically flat as $\bar{r} \rightarrow + \infty$,
and singular on the hypersurface
 $\bar{r} = 0$. This can be seen, for example,
from the Kretschmann scalar,
\bq
\lb{0.1}
R_{\mu\nu\beta\gamma} R^{\mu\nu\beta\gamma}
= \frac{(n^{2} + 3)(n^{2} - 1)^{2}}{4\bar{r}^{n^{2} + 3}}.
\eq
Thus, the Kretschmann scalar diverges as $\bar{r} \rightarrow 0$,
except for the cases $n = \pm 1$. In the last two cases, it can be
shown that the spacetimes are flat and belong to Petrov type $O$. When 
$n = 0, \; 3$, the spacetimes are Petrov type $D$, while all the rest
are Petrov type $I$. 
 
In order to apply the formulas developed in the last section, we need
first to write the Lewis solutions in the form of Eq.(\ref{eq1}). To this
end, we make the following coordinate transformations
\bqn
\lb{4.4}
\bar{t}&=& \left\{\frac{\bar{\alpha}}{\sqrt{a}}
\left(1 - \frac{bc}{n}\right) - \frac{b\gamma \delta}{R_{0}^{(1+n)/2}}
\right\} t - \frac{b\gamma}{R_{0}^{(1+n)/2}}\varphi,\nb\\
\bar{\varphi}&=& \left\{\frac{\bar{\alpha}c}{n\sqrt{a}}
 + \frac{\gamma \delta}{R_{0}^{(1+n)/2}}
\right\} t + \frac{\gamma}{R_{0}^{(1+n)/2}}\varphi,\nb\\
\bar{r}&=& R_{0}(r + d)^{4/(1+n)^{2}}, \nb\\
\bar{z} &=& \frac{\beta z}{ R_{0}^{(n^{2} - 1)/4}},\;\; (n \not= -1),
\eqn
for $n \not= -1$, where $\bar{\alpha},\; \beta,\; \gamma,\;
\delta$ and $d$ are arbitrary constants, and $R_{0} \equiv
[\bar{\alpha}(1+n)^{2}/4]^{4/(1+n)^{2}}$. Then, the Lewis solutions
takes the form,
\bq
\lb{4.5}
ds^{2}_{+} = \alpha^{2}A^{4\sigma}(r)(dt^{2} - dr^{2})
- \beta^{2}A^{4\sigma(2\sigma-1)}(r)dz^{2}
- \frac{\gamma^{2}}{a}A^{2(1-2\sigma)}(r)(d\varphi + \delta dt)^{2},
\;\;\; (\sigma \not= 1/2),
\eq
where
\bq
\lb{4.6}
A(r) \equiv (r + d)^{1/(2\sigma -1)^{2}},\;\;\;
\alpha \equiv \bar{\alpha}R_{0}^{2\sigma},\;\;\;
\sigma \equiv \frac{1 - n}{4}.
\eq
From the above expressions we can see that the range of $\bar{r}$,
$\bar{r} \in [0 ,\; \infty)$, is mapped into the range $r \in [-d,\;
\infty)$. Hence, when $r \rightarrow + \infty$,
the spacetime is asymptotically flat, and when $r \rightarrow - d$
it is singular, except for the case $n = 1$ where the spacetime
is flat. As we shall see blow, we will use the above metric
as describing the spacetime outside of a stationary shell. Thus,
to prevent spacetime singularity from happening outside the shell, 
we shall require that $r + d > 0$. 

When $\sigma = 1/2 \;$ (or $n = -1$), the coordinate transformations
\bqn
\lb{4.7}
\bar{t}&=& \left\{\frac{\bar{\alpha}}{\sqrt{a}}
(1 + bc) - b\gamma \delta
\right\} t - b\gamma\varphi,\nb\\
\bar{\varphi}&=& - \left\{\frac{\bar{\alpha}c}{\sqrt{a}}
  - \gamma \delta \right\} t + \gamma\varphi,\nb\\
\bar{r}&=& e^{\alpha (r + d)}, \nb\\
\bar{z} &=& z,\;\;\; (\sigma = 1/2),
\eqn
bring the metric (\ref{4.2}) to the form
\bq
\lb{4.8}
ds^{2}_{+} = \alpha^{2}e^{2\alpha (r + d)}(dt^{2} - dr^{2})
- dz^{2}
- \frac{\gamma^{2}}{a}(d\varphi + \delta dt)^{2},
\;\;\; (\sigma = 1/2).
\eq
As shown above, this metric represents a flat spacetime but in a
rotating coordinate system. 
%As a result, the
%coordinate $r$ can be taken any real value that belongs to 
%[$- d,\; +\infty$).

It should be noted that the above coordinate transformations of
Eqs.(\ref{4.4}) and (\ref{4.7}) are 
admissible only locally,
{\em if both sets of the coordinates are considered as representing 
 cylindrical coordinates}. Otherwise,
it will give rise to new topology 
 \cite{MS1998,MacCullum1997}. However, in this paper
we shall take the point of view that only the coordinates 
$\{x^{\mu}\} = \{t,\; r,\; z,\; \varphi\}$ represent 
the usual cylindrical coordinates. Then, the topological 
identifications discussed in  \cite{MS1998,MacCullum1997}
are not applicable to the present case, so that the above
coordinate transformations are admissible even globally.

Assuming that a stationary shell located on the hypersurface
$r = r_{0} > 0$ generates the spacetime 
described by the metric Eq.(\ref{4.5}) 
or Eq.(\ref{4.8}), and that the spacetime inside the shell is vacuum
and described by the metric  Eq.(\ref{4.1}), where $r_{0}$ is a constant,
we find that the first junction conditions 
that the metric are continuous across
the surface, i.e., 
\bq
\lb{4.9a}
g^{-}_{\mu\nu}(r\rightarrow r_{0}^{-})
= g^{+}_{\mu\nu}(r\rightarrow r_{0}^{+}), 
\eq
lead to  
\bq
\lb{4.9}
\alpha = A^{- 2\sigma}(r_{0}),\;\; 
\beta = A^{- 2 \sigma(2\sigma -1 )}(r_{0}),\;\;
\gamma = \sqrt{a}r_{0}A^{2\sigma -1}(r_{0}),\;\;
\delta = \Omega, \;\; (\sigma \not= 1/2),
\eq
for $\sigma \not= 1/2$, and 
\bq
\lb{4.10}
\alpha e^{\alpha(r_{0}+d)}= 1,\;\; 
\gamma = \sqrt{a}r_{0},\;\;
\delta = \Omega, \;\; (\sigma = 1/2),
\eq
for $\sigma = 1/2$.
Substituting Eqs.(\ref{4.9}) and (\ref{4.10}), respectively, into
Eq.(\ref{4.5}) and Eq.(\ref{4.8}), we find that
\bq
\lb{4.11}
ds^{2}_{+} = B^{4\sigma}(r)(dt^{2} - dr^{2})
- B^{4\sigma(2\sigma-1)}(r)dz^{2}
- r_{0}^{2}B^{2(1-2\sigma)}(r)(d\varphi + \Omega dt)^{2},
\;\;\; (\sigma \not= 1/2),
\eq
for $\sigma \not= 1/2$, where 
\bq
\lb{4.12}
B(r) \equiv \left(\frac{r +d}{r_{0} + d}\right)^{1/(2\sigma - 1)^{2}},
\; (\sigma \not= 1/2),
\eq
and
\bq
\lb{4.13}
ds^{2}_{+} = e^{2\alpha (r - r_{0})}(dt^{2} - dr^{2})
- dz^{2}
- r^{2}_{0}(d\varphi + \Omega dt)^{2},
\;\;\; (\sigma = 1/2),
\eq
for $\sigma = 1/2$.

Now we are at the position of calculating the surface
EMT of the shell. In the following, let us consider the two cases,
$\sigma \not= 1/2$ and $\sigma = 1/2$, separately.

Case a) {$\;\sigma \not= 1/2 $}:
In this case, from Eqs.(\ref{eq1}), (\ref{4.1}), (\ref{4.12}) 
and (\ref{eq19}), we find that
\bqn
\lb{4.14}
\left[\psi_{,r}\right]^{-} & =&
\frac{2\sigma}{(2\sigma -1)(r_{0} + d)},\nb\\
\left[\chi_{,r}\right]^{-} & =& 
\frac{4\sigma^{2}}{(2\sigma -1)^{2}(r_{0} + d)},\nb\\
\left[\omega_{,r}\right]^{-} & =& 0,\nb\\
\left[W_{,r}\right]^{-} & =& -\frac{d}{r_{0} + d},\;\; 
(\sigma \not = 1/2).
\eqn
Substituting the above expressions into Eq.(\ref{eq25}), we find
\bqn
\lb{4.15}
\eta &=&  \frac{d}{r_{0}(r_{0} + d)},\nb\\ 
q  &=& 0, \nb\\      
p_{z} & =&  \frac{4\sigma(1 - \sigma)r_{0} - (2\sigma - 1)^{2} d}
{(2\sigma - 1)^{2}r_{0}(r_{0} + d)},\nb\\  
p_{\varphi} &=& \frac{4\sigma^{2}}
{(2\sigma - 1)^{2}(r_{0} + d)},\;\; (\sigma  \not= 1/2).  
\eqn
The combination of Eqs.(\ref{4.15}) and (\ref{eq27d}) shows that in the
present case we have $D > 0$. Thus, now the fluid is type $I$.

For the shell to be physically acceptable, it has to satisfy some   energy
conditions, weak, dominant, or strong \cite{HE1973}. It can be shown that if
the condition \bq \lb{4.16}
0 \le \sigma \le 1,\;\;\;\; d \ge 0,
\eq
holds, the weak energy condition will be satisfied, and
if the condition
\bq
\lb{4.17}
d \ge \cases{  
\frac{2\sigma(1 - \sigma)r_{0}}{(2\sigma - 1)^{2}},
& $0 \le \sigma \le \frac{1}{3}$,\cr
 \frac{4\sigma^{2}r_{0}}{(2\sigma - 1)^{2}}, &
$\frac{1}{3} < \sigma \le 1$, \cr}
\eq
holds, the dominant energy condition will be satisfied,
while if the condition 
\bq
\lb{4.18}
d \ge \cases{
- \frac{4\sigma^{2}r_{0}}{(2\sigma - 1)^{2}},
& $0 \le \sigma \le \frac{1}{4}$,\cr
 - r_{0},  &
$\frac{1}{4} < \sigma \le 1$, 
\cr}.
\eq
holds, the strong energy condition will be satisfied.
Clearly, for $ 0 \le \sigma \le 1$, all the
three energy conditions can be satisfied by properly choosing
the free parameter $d$. 
It is interesting to note that in the case
$\Omega = 0$ it was shown that the solutions have physics only
when $ 0 \le \sigma \le 1/2$ \cite{Bonnor1992}. Recently
we extended  it to the range $\sigma \in [0,\; 1]$ \cite{WSS1997}, 
which is exactly the same as that obtained above for  the case with 
rotation, $\Omega \not= 0$. 

Case b) {$\;\sigma  = 1/2$}:
In this case, Combining  Eqs.(\ref{eq1}), (\ref{4.1}), (\ref{4.13}) 
and (\ref{eq19}), we find that
\bq
\lb{4.19}
\left[\psi_{,r}\right]^{-} = 0,\;\;\;
\left[\chi_{,r}\right]^{-} = \alpha,\;\;\;
\left[\omega_{,r}\right]^{-} = 0,\;\;\;
\left[W_{,r}\right]^{-}  = - 1,\;\; (\sigma = 1/2).
\eq
Then, the surface EMT of the shell is given by Eq.(\ref{eq24})
with  
\bqn
\lb{4.20}
\eta &=&  \frac{1}{r_{0}} ,\nb\\ 
q  &=& 0, \nb\\    
p_{z} & =&  \frac{\alpha r_{0} - 1 }{r_{0}} ,\nb\\  
p_{\varphi} &=&  \alpha,\;\; (\sigma  = 1/2).   
\eqn
It can be shown that when $\alpha \ge 0$ the weak and strong energy
conditions will be satisfied, while when $0 \le \alpha \le 1/r_{0}$
the dominant energy condition will be satisfied. Similar to the last case, the
fluid now is also type $I$.

\section{Conclusions}       

In this paper, the main properties of cylindrical spacetimes with rotation
were studied by using the NP formulas. The physical interpretations
of each component of the Riemann tensor was given by considering 
null geodesic deviations. It would be very interesting to see their
experimental implications.

In Sec. $III$ the spacetimes were extended to include
rotating cylindrical shells, and the general expression of the surface
energy-momentum tensor of the shells were given in terms of the 
discontinuation of the first derivatives of the metric coefficients.
This would be very useful in studying gravitational 
collapse of a rotating cylindrical shell. As a matter of fact, this was
one of our main motivations for such a study. As an application
of the formulas developed, we considered a stationary shell that generates
the Lewis solutions, which represent the most general vacuum solutions
of the cylindrical spacetimes with rotation \cite{Kramer1980}, 
by assuming that inside the 
shell the spacetime is flat. It was shown that by properly choosing one of
the free parameters, the shell can satisfy all the three energy conditions,
provided $ 0 \le \sigma \le 1$, where $\sigma $ is related to the mass
per unit length of the cylindrical shell \cite{Bonnor1992,Fatima1995}. 
This range is exactly the same
as that obtained in the static case \cite{WSS1997}.

%%%%%%%%%%%%%%%%%%%%%%%%%%%%%%%%%%%%%%%%%%%%%%%%%%%%%%%%%%%%%%%%%%%%%%%%
%\newpage
\section*{Appendix {A}: The Christoffel Symbols and the Einstein
      Tensor}

\renewcommand{\theequation}{A.\arabic{equation}}
\setcounter{equation}{0}

Corresponding to the metric (\ref{eq1}), the non-vanishing Christoffel
symbols, defined by
\bq
\lb{b0}
\Gamma^{\mu}_{\nu\lambda} = \frac{1}{2}g^{\mu\sigma}\left\{
        g_{\sigma\lambda,\nu} + g_{\nu\sigma,\lambda}
        - g_{\nu\lambda,\sigma}\right\},
\eq
are given by
\bqn
\lb{b1}
\Gamma^{0}_{00} &=& \chi_{,t} - \psi_{,t} - \omega^{2}W^{2}e^{-2\chi} 
                    \left(\psi_{,t} - \frac{W_{,t}}{W}\right),\;\;
\Gamma^{0}_{01} = \chi_{,r} - \psi_{,r} 
                  - \frac{1}{2}\omega W^{2}e^{-2\chi} \omega_{,r},\nb\\
\Gamma^{0}_{03} &=&  - \omega W^{2}e^{-2\chi}\left(\psi_{,t} 
                     - \frac{W_{,t}}{W}\right),\;\;
\Gamma^{0}_{11} = \chi_{,t} - \psi_{,t},\;\;
\Gamma^{0}_{13} = - \frac{1}{2}W^{2}e^{-2\chi} \omega_{,r},\nb\\
\Gamma^{0}_{22} &=& e^{2(2\psi - \chi)} \psi_{,t},\;\;
\Gamma^{0}_{33} = - W^{2}e^{-2\chi}\left(\psi_{,t} - 
                  \frac{W_{,t}}{W}\right),\nb\\
\Gamma^{1}_{00} &=& \chi_{,r} - \psi_{,r} + \omega^{2}W^{2}e^{-2\chi} 
                    \left(\psi_{,r} - \frac{W_{,r}}{W}-
                     \frac{\omega_{,r}}{\omega}\right),\;\;
\Gamma^{1}_{01} = \chi_{,t} - \psi_{,t},\nb\\
\Gamma^{1}_{03} & =& \frac{1}{2}\omega W^{2}e^{-2\chi}
                 \left(2\psi_{,r} - 2\frac{W_{,r}}{W}
                 - \frac{\omega_{,r}}{\omega}\right),\;\;
\Gamma^{1}_{11} = \chi_{,r} - \psi_{,r},\nb\\
\Gamma^{1}_{22} &=& - e^{2(2\psi - \chi)} \psi_{,r},\;\;
\Gamma^{1}_{33} = W^{2}e^{-2\chi}\left(\psi_{,r} - 
                  \frac{W_{,r}}{W}\right),\;\;
\Gamma^{2}_{02} = \psi_{,t},\;\;
\Gamma^{2}_{12} = \psi_{,r},\nb\\
\Gamma^{3}_{00} &=& - \omega\left(\chi_{,t} + \psi_{,t} - 2\frac{W_{,t}}{W}
                    - \frac{\omega_{,r}}{\omega} \right)
                    + \frac{1}{2}\omega^{2}W^{2}e^{-2\chi} \omega_{,r},\nb\\
\Gamma^{3}_{03} &=& - (1 - \omega^{2}W^{2}e^{-2\chi})\left[\psi_{,t} - 
                  \frac{W_{,t}}{W}\right],\;\;
\Gamma^{3}_{11} = - \omega(\chi_{,t} - \psi_{,t}),\nb\\    
\Gamma^{3}_{13} &=& \frac{W_{,r}}{W} - \psi_{,r} 
                    +\frac{1}{2}\omega W^{2}e^{-2\chi} \omega_{,r},\;\;
\Gamma^{3}_{22} = - \omega e^{2(2\psi - \chi)} \psi_{,t},\nb\\
\Gamma^{3}_{33} &=& \omega W^{2}e^{-2\chi}\left[\psi_{,t} - 
                  \frac{W_{,t}}{W}\right],                                      
 \eqn
while the non-vanishing components of the Einstein tensor are given
by
\bqn
\lb{b2}
G_{00} &=&  - \left\{\frac{W_{,rr}}{W} - \frac{1}{W}(\chi_{,t}W_{,t}
           + \psi_{,r}W_{,r}) + \psi_{,t}^{2} + \psi_{,r}^{2}
           - \frac{3}{4}\omega^{2}W^{4}e^{-4\chi}\omega_{,r}^{2}\right.\nb\\
          & & \left. + \omega^{2}W^{2}e^{-2\chi}\left[\chi_{,tt} - \chi_{,rr}
          + \frac{\omega_{,rr}}{\omega} + \psi_{,t}^{2} - \psi_{,r}^{2} 
          - 2\frac{\chi_{,r}\omega_{,r}}{\omega} 
          + 3\frac{\omega_{,r}W_{,r}}{\omega W} 
          + \frac{1}{4}\left(\frac{\omega_{,r}}{\omega}
          \right)^{2}\right]\right\},\nb\\
G_{01} &=&  - \left\{\frac{W_{,tr}}{W} - \frac{1}{W}(\chi_{,t}W_{,r}
           + \chi_{,r}W_{,t}) + 2\psi_{,t} \psi_{,r}\right.\nb\\
           & & \left. 
           + \frac{1}{2}\omega W^{2}e^{-2\chi}
           \left[\omega_{,tr} - 2\chi_{,t}\omega_{,r} +  
          + 3\omega_{,r}\frac{W_{,t}}{W} \right]\right\},\nb\\          
G_{03} &=&  - \omega W^{2}e^{-2\chi}\left\{
           \chi_{,tt} - \chi_{,rr} + \frac{\omega_{,rr}}{2\omega} 
           + \psi_{,t}^{2} - \psi_{,r}^{2} 
           - \frac{\chi_{,r}\omega_{,r}}{\omega}
           + \frac{3\omega_{,r}W_{,r}}{2\omega W}
           + \frac{3}{4}W^{2}e^{-2\chi}\omega^{2}_{,r} 
           \right\},\nb\\   
G_{11} &=&  - \left\{\frac{W_{,tt}}{W}  + \psi_{,t}^{2} + \psi_{,r}^{2} 
           - \frac{1}{W}(\chi_{,t}W_{,t} + \chi_{,r}W_{,r})
           - \frac{1}{4}W^{2}e^{-2\chi}\omega^{2}_{,r} 
           \right\},\nb\\ 
G_{13} &=&  - \frac{1}{2}W^{2}e^{-2\chi}\left\{
               \omega_{,tr} + 3\omega_{,r}\frac{W_{,t}}{W} 
               - 2\chi_{,t}\omega_{,r} 
              \right\},\nb\\
G_{22} &=&  e^{2(2\psi - \chi)} \left\{
        2(\psi_{,tt} - \psi_{,rr}) - (\chi_{,tt} - \chi_{,rr})
        - \frac{1}{W}(W_{,tt} - W_{,rr})\right. \nb\\
        & &  \left. - (\psi_{,t}^{2} - \psi_{,r}^{2})
        + \frac{2}{W}(\psi_{,t}W_{,t} - \psi_{,r}W_{,r})
        - \frac{1}{4}W^{2}e^{-2\chi}\omega_{,r}^{2}
         \right\},\nb\\
G_{33} &=&  - W^{2}e^{- 2\chi} \left\{
         \chi_{,tt} - \chi_{,rr}
        + \psi_{,t}^{2} - \psi_{,r}^{2}
        + \frac{3}{4}W^{2}e^{-2\chi}\omega_{,r}^{2}
         \right\},  
\eqn  
where $\chi_{,r}\equiv {\partial\chi}/{\partial r}$ and 
$\chi_{,t}\equiv {\partial \chi}/{\partial t}$, etc.
      
\section*{Appendix {B}: The Spin coefficients and the Ricci and 
              Weyl scalars}

\renewcommand{\theequation}{B.\arabic{equation}}
\setcounter{equation}{0}

Choosing the null tetrad as given by Eq.(\ref{eq2}), 
we find that the 
spin coefficients are given by
\bqn
\lb{a1}
\rho&=&l_{\mu;\nu}m^{\mu}\bar m^{\nu}=
-\frac{e^{\psi -\chi}}{2\sqrt {2} A W}({W_{,t}}
+ {W_{,r}}), \nb\\
\kappa&=&l_{\mu;\nu}m^{\mu}l^{\nu}=0 \nb\\
\sigma&=&l_{\mu;\nu}m^{\mu}m^{\nu}=
-\frac{e^{\psi -\chi}}{2\sqrt{2}A}\left[2(\psi_{,t}+
\psi_{,r}) - \frac{W_{,t}+{W_{,r}}}{W}\right], \nb\\
\tau&=&l_{\mu;\nu}m^{\mu}n^{\nu}=-i \frac{W e^{\psi - 2\chi}}
{2\sqrt 2}\omega_{,r}, \nb\\
\alpha&=&\frac{1}{2}\left(l_{\mu;\nu}n^{\mu}\bar m^{\nu}
-m_{\mu;\nu}\bar m^{\mu}\bar m^{\nu}\right)=
i \frac{W e^{\psi - 2\chi}}{4\sqrt 2}\omega_{,r}, \nb\\
\epsilon&=&\frac{1}{2}\left(l_{\mu;\nu}n^{\mu}l^{\nu}-m_{\mu;\nu}
\bar m^{\mu}l^{\nu}\right)= \frac{e^{\psi -\chi}}
{2\sqrt{2} A}\left[\chi_{,t}+\chi_{,r}-(\psi_{,t}+\psi_{,r})
- \frac{A_{,t} + A_{,r}}{A}\right],\nb\\
\mu&=&-n_{\mu;\nu}\bar{m}^{\mu}m^{\nu}=
\frac{A e^{\psi -\chi}}{2\sqrt {2}W}
({W_{,t}}-{W_{,r}}), \nb\\
\nu&=&-n_{\mu;\nu}\bar{m}^{\mu}n^{\nu}=0, \nb\\
\lambda&=&-n_{\mu;\nu}\bar m^{\mu}\bar m^{\nu}=
\frac{A e^{\psi -\chi}}{2\sqrt {2}}
\left[2(\psi_{,t}-\psi_{,r})-\frac{{W_{,t}} - {W_{,r}}}{W}\right],
\nb\\
\pi&=&-n_{\mu;\nu}\bar{m}^{\mu}l^{\nu}=
i\frac{W e^{\psi - 2\chi}}{2\sqrt 2}\omega_{,r}, \nb\\
\beta&=&\frac{1}{2}\left(l_{\mu;\nu}n^{\mu}m^{\nu}-
m_{\mu;\nu}\bar m^{\mu}m^{\nu}\right)=
-i\frac{W e^{\psi - 2\chi}}{4\sqrt 2}\omega_{,r}, \nb\\
\gamma&=&\frac{1}{2}\left(l_{\mu;\nu}n^{\mu}n^{\nu}-m_{\mu;\nu}
\bar m^{\mu}n^{\nu}\right)= - \frac{A e^{\psi -\chi}}
{2\sqrt 2}\left[\chi_{,t}- \chi_{,r} - (\psi_{,t}-\psi_{,r})
+ \frac{A_{,t} - A_{,r}}{A}\right].
\eqn 
Then, the corresponding Ricci and Weyl 
scalars are given, respectively, by
\begin{eqnarray}
\lb{a2}
\Phi_{00}&=&  \frac{1}{2}S_{\mu\nu}l^{\mu}l^{\nu}
           \nonumber \\
         &=& - \frac{e^{2(\psi - \chi)}}{4A^{2}W}\left[W_{,tt}
         + 2W_{,tr} + W_{,rr} 
         - 2(\chi_{,t}+\chi_{,r})(W_{,t}+W_{,r})
         + 2W (\psi_{,t} + \psi_{,r})^{2}\right],
          \nb\\
\Phi_{01}&=& \frac{1}{2}S_{\mu\nu}l^{\mu}m^{\nu} 
            \nonumber \\
         &=& - i \frac{W e^{2\psi - 3\chi}}{8A}\left[\omega_{,rr}
         + \omega_{,tr} 
         - 2\omega_{,r}(\chi_{,t}+\chi_{,r}) 
         + \frac{3\omega_{,r}}{W}(W_{,t} + W_{,r})\right],
          \nb\\
\Phi_{02}&=&\frac{1}{2}S_{\mu\nu} m^{\mu} m^{\nu}  \nonumber \\
         &=& \frac{1}{2}e^{2(\psi - \chi)}\left[\psi_{,tt}
         -\psi_{,rr} + \frac{1}{W} (\psi_{,t}W_{,t} -
         \psi_{,r}W_{,r})
         - \frac{1}{2W}(W_{,tt} - W_{,rr}) +
         \frac{1}{4}W^{2}e^{-2\chi}\omega^{2}_{,r}\right],
          \nb\\
\Phi_{11}&=&\frac{1}{4}S_{\mu\nu}(l^{\mu}n^{\nu}+m^{\mu}\bar m^{\nu})
           \nonumber \\
         &=& \frac{1}{4}e^{2(\psi-\chi)}\left[
         \psi_{,tt} - \psi_{,rr} - \psi^{2}_{,t} + \psi_{,r}^{2}
         + \frac{1}{W}(\psi_{,t}W_{,t} - \psi_{,r}W_{,r} )
         + \chi_{,rr} - \chi_{,tt} - \frac{3}{4}W^{2}e^{-2\chi}
         \omega_{,r}^{2}\right],\nb\\
\Phi_{12}&=&\frac{1}{2}S_{\mu\nu}n^{\mu}m^{\nu}  \nonumber \\
         &=& - i \frac{1}{8}A W e^{2\psi - 3\chi}\left[\omega_{,rr}
         -\omega_{,tr} + 2\omega_{,r}(\chi_{,t} - \chi_{,r})
         - \frac{3\omega_{,r}}{W}(W_{,t}- W_{,r})\right],
          \nb\\ 
\Phi_{22}&=&\frac{1}{2}S_{\mu\nu}n^{\mu}n^{\nu}  \nonumber \\       
         &=& - \frac{A^{2}e^{2(\psi - \chi)}}{4W}\left[W_{,tt}
         - 2W_{,tr} + W_{,rr} 
         - 2(\chi_{,t}-\chi_{,r})(W_{,t}-W_{,r})
         + 2W (\psi_{,t} - \psi_{,r})^{2}\right],
         \nb\\
\Lambda&=&\frac{1}{24}R   \nonumber \\
       &=& - \frac{1}{12}e^{2(\psi - \chi)}\left[\psi_{,tt}
         - \psi_{,rr} - \psi^{2}_{,t} + \psi^{2}_{,r} +
         \frac{1}{W}(\psi_{,t}W_{,t} - \psi_{,r}W_{,r})
         - (\chi_{,tt} - \chi_{,rr}) \right.\nb\\
         & & \left.- \frac{1}{W}(W_{,tt} - 
         W_{,rr}) n- \frac{1}{4}W^{2}e^{-2\chi}\omega^{2}_{,r}\right],
\eqn
and
\bqn
\lb{a3}          
\Psi_{0}&=&-C_{\mu\nu\lambda\delta}l^{\mu}m^{\nu}l^{\lambda}m^{\delta}
            \nonumber \\
          &=& - \frac{e^{2(\psi-\chi)}}{2A^{2}}\left\{
         \psi_{,tt} + 2 \psi_{,tr} + \psi_{,rr} + 2(\psi_{,t} + \psi_{,r})^{2}
         - 2(\psi_{,t}+\psi_{,r})(\chi_{,t} + \chi_{,r})\right.\nb\\
         & &- \left.\frac{1}{2W}\left[W_{,tt} + 2W_{,tr} + W_{,rr}
         - 2(W_{,t} + W_{,r})(\chi_{,t} + \chi_{,r})
         \right]\right\},\nb\\
\Psi_{1}&=&-C_{\mu\nu\lambda\delta}l^{\mu}n^{\nu}l^{\lambda}m^{\delta}
           \nonumber \\
        &=&  - i \frac{W e^{2\psi - 3\chi}}{8A}\left[\omega_{,rr}
         +\omega_{,tr} - 2\omega_{,r}(\psi_{,t} + \psi_{,r} + \chi_{,t} 
         + \chi_{,r})  
         + \frac{3\omega_{,r}}{W}(W_{,t}+ W_{,r})\right],
          \nb\\ 
\Psi_{2}&=&-\frac{1}{2}C_{\mu\nu\lambda\delta}[l^{\mu}n^{\nu}
           l^{\lambda}n^{\delta}-l^{\mu}n^{\nu}
            m^{\lambda}\bar m^{\delta}  ]\nonumber \\
          &=& \frac{1}{6}e^{2(\psi - \chi)}
          \left\{\psi_{,tt}
         - \psi_{,rr} - \chi_{,tt} + \chi_{,rr} +
         2(\psi^{2}_{,t} - \psi^{2}_{,r}) \right.\nb\\
         & & +\left.
         \frac{1}{2W}\left[W_{,tt} - W_{,rr} - 4(\psi_{,t}W_{,t}
         - \psi_{,r}W_{,r})\right]
         - W^{2}e^{-2\chi} \omega^{2}_{,r}\right\},
         \nb\\
\Psi_{3}&=& - C_{\mu\nu\lambda\delta}n^{\mu}
         l^{\nu}n^{\lambda}\bar m^{\delta}   \nonumber \\
        &=&  i \frac{1}{8}A W e^{2\psi - 3\chi}\left[\omega_{,rr}
         - \omega_{,tr} + 2\omega_{,r}(\psi_{,t} - \psi_{,r} + \chi_{,t} 
         - \chi_{,r})  
         - \frac{3\omega_{,r}}{W}(W_{,t}- W_{,r})\right],
          \nb\\ 
\Psi_{4}&=&-C_{\mu\nu\lambda\delta}n^{\mu}\bar m^{\nu}
           n^{\lambda}\bar m^{\delta} 
           \nb\\
           &=&- \frac{1}{2}A^{2}e^{2(\psi-\chi)}\left\{
         \psi_{,tt} - 2 \psi_{,tr} + \psi_{,rr} + 2(\psi_{,t} - \psi_{,r})^{2}
         - 2(\psi_{,t}-\psi_{,r})(\chi_{,t} - \chi_{,r})\right.\nb\\
         & &\left.- \frac{1}{2W}\left[W_{,tt} - 2W_{,tr} + W_{,rr}
         - 2(W_{,t} - W_{,r})(\chi_{,t} - \chi_{,r})\right]\right\},       
\eqn
where 
\bq
\lb{a4}
S_{\mu\nu} \equiv R_{\mu\nu} - \frac{1}{4}g_{\mu\nu}R,
\eq
and $C_{\mu\nu\sigma\delta}$ denotes the Weyl tensor. In terms of
the Riemann and Ricci tensors, it is given by 
\bqn
\lb{a5}
C_{\mu\nu\lambda\delta} &=& R_{\mu\nu\lambda\delta}
- \frac{1}{2}\left(g_{\mu\lambda}R_{\nu\delta} 
+ g_{\nu\delta}R_{\mu\lambda} 
- g_{\nu\lambda}R_{\mu\delta}
- g_{\mu\delta}R_{\nu\lambda}\right)\nb\\
& & - \frac{1}{6}\left(g_{\mu\delta}g_{\nu\lambda}
- g_{\mu\lambda}g_{\nu\delta}\right)R.
\eqn

In terms of the Weyl and Ricci scalars, the Weyl and Ricci tensors are
given, respectively, by 
\bqn
\lb{a6}
C_{\mu\nu\lambda\delta} &=& - 4\left\{
        \left(\Psi_{2} + \bar{\Psi}_{2}\right)
    \left(l_{[\mu}n_{\nu]}l_{[\lambda}n_{\delta]}
    +  m_{[\mu}\bar{m}_{\nu]}m_{[\lambda}\bar{m}_{\delta]}\right)
    \right.\nb\\
    & &
    - \left(\Psi_{2} - \bar{\Psi}_{2}\right)
    \left(l_{[\mu}n_{\nu]}m_{[\lambda}\bar{m}_{\delta]} 
    + m_{[\mu}\bar{m}_{\nu]}l_{[\lambda}n_{\delta]}\right)
    \nb\\
    & & + \Psi_{0}n_{[\mu}\bar{m}_{\nu]}n_{[\lambda}\bar{m}_{\delta]}\nb\\
    & &
    + \Psi_{1}\left(l_{[\mu}n_{\nu]}n_{[\lambda}\bar{m}_{\delta]}
     + n_{[\mu}\bar{m}_{\nu]}l_{[\lambda}n_{\delta]}
     + n_{[\mu}\bar{m}_{\nu]}\bar{m}_{[\lambda}m_{\delta]}
     + \bar{m}_{[\mu}{m}_{\nu]}n_{[\lambda}\bar{m}_{\delta]}\right)
     \nb\\
   & & - \Psi_{2}\left(l_{[\mu}{m}_{\nu]}n_{[\lambda}\bar{m}_{\delta]}
       + n_{[\mu}\bar{m}_{\nu]}l_{[\lambda}m_{\delta]}\right)\nb\\
   & & - \Psi_{3}\left(l_{[\mu}n_{\nu]}l_{[\lambda}m_{\delta]}
        + l_{[\mu}{m}_{\nu]}l_{[\lambda}n_{\delta]}
        - l_{[\mu}{m}_{\nu]}m_{[\lambda}\bar{m}_{\delta]}
        - m_{[\mu}\bar{m}_{\nu]}l_{[\lambda}m_{\delta]}\right)\nb\\
   & & + \Psi_{4}l_{[\mu}{m}_{\nu]}l_{[\lambda}m_{\delta]}\nb\\
   & & + \bar{\Psi}_{0}n_{[\mu}{m}_{\nu]}n_{[\lambda}{m}_{\delta]}\nb\\
   & &
    + \bar{\Psi_{1}}\left(l_{[\mu}n_{\nu]}n_{[\lambda}{m}_{\delta]}
     + n_{[\mu}{m}_{\nu]}l_{[\lambda}n_{\delta]}
     + n_{[\mu}{m}_{\nu]}{m}_{[\lambda}\bar{m}_{\delta]}
     + {m}_{[\mu}\bar{m}_{\nu]}n_{[\lambda}{m}_{\delta]}\right)
     \nb\\
   & & - \bar{\Psi}_{2}\left(l_{[\mu}\bar{m}_{\nu]}n_{[\lambda}{m}_{\delta]}
       + n_{[\mu}{m}_{\nu]}l_{[\lambda}\bar{m}_{\delta]}\right)\nb\\
   & & - \bar{\Psi}_{3}\left(l_{[\mu}n_{\nu]}l_{[\lambda}\bar{m}_{\delta]}
        + l_{[\mu}\bar{m}_{\nu]}l_{[\lambda}n_{\delta]}
        - l_{[\mu}\bar{m}_{\nu]}\bar{m}_{[\lambda}{m}_{\delta]}
        - \bar{m}_{[\mu}{m}_{\nu]}l_{[\lambda}\bar{m}_{\delta]}\right)\nb\\
& & \left. 
   + \Psi_{4}l_{[\mu}\bar{m}_{\nu]}l_{[\lambda}\bar{m}_{\delta]}\right\},
\eqn
and
\bqn
\lb{a7}
R_{\mu\nu} &=& 2\left\{
        \Phi_{00}n_{\mu}n_{\nu} + \Phi_{22}l_{\mu}l_{\nu}\right.
             \nb\\
        & &   - \Phi_{01}\left(n_{\mu}\bar{m}_{\nu} 
             + n_{\nu}\bar{m}_{\mu}\right)
              - \bar{\Phi}_{01}\left(n_{\mu}{m}_{\nu} 
             + n_{\nu}{m}_{\mu}\right)\nb\\
       & &  + \Phi_{02}\bar{m}_{\mu}\bar{m}_{\nu} 
             + \bar{\Phi}_{02}m_{\nu}{m}_{\mu}\nb\\
       & & \left(2 \Phi_{11}- 3\Lambda\right)
            \left(l_{\mu}n_{\nu} 
             + l_{\nu}n_{\mu}\right)
             + \left(2 \Phi_{11} + 3\Lambda\right)
             \left(m_{\mu}\bar{m}_{\nu} 
             + m_{\nu}\bar{m}_{\mu}\right)\nb\\
       & & \left.- \Phi_{12}\left(l_{\mu}\bar{m}_{\nu} 
             + l_{\nu}\bar{m}_{\mu}\right)
              - \bar{\Phi}_{12}\left(l_{\mu}{m}_{\nu} 
             + l_{\nu}{m}_{\mu}\right)\right\}.
\eqn             
It should be noted that Eqs.(\ref{a6}) and (\ref{a7}) hold for
the general case.

\section*{Acknowledgment} 

We would like to thank P.S. Letelier, N.O. Santos, and M.F.A. da Silva for
useful discussions.  The financial assistance from  CNPq and FAPERJ (AW) is
gratefully acknowledged.     
%%%%%%%%%%%%%%%%%%%%%%%%%%%%%%%%%%%%%%%%%%%%%%%%%%%%%%%%%%%%%%% 
%\newpage

%%%%%%%%%%%%%%%%%%%%%%%%%%%%%%%%%%%%%%%%%%%%%%%%%%%%%%%%%%%%%%%%%%%%%%
\newpage

\section*{FIGURE CAPTIONS}

{\bf Fig.1} $S_{O}$ and $S_{P}$ are two infinitesimal 2-elements 
spanned by $e_{2}$
and $e_{3}$ and orthogonal to the null 
geodesic $C$ defined by $l^{\mu}$,
passing $S_{O}$ and $S_{P}$ at the points $O$ and $P$, respectively. 
A light beam meets $S_{O}$ in the circle $S$.
\vspace{1 cm}

\centerline{\epsffile{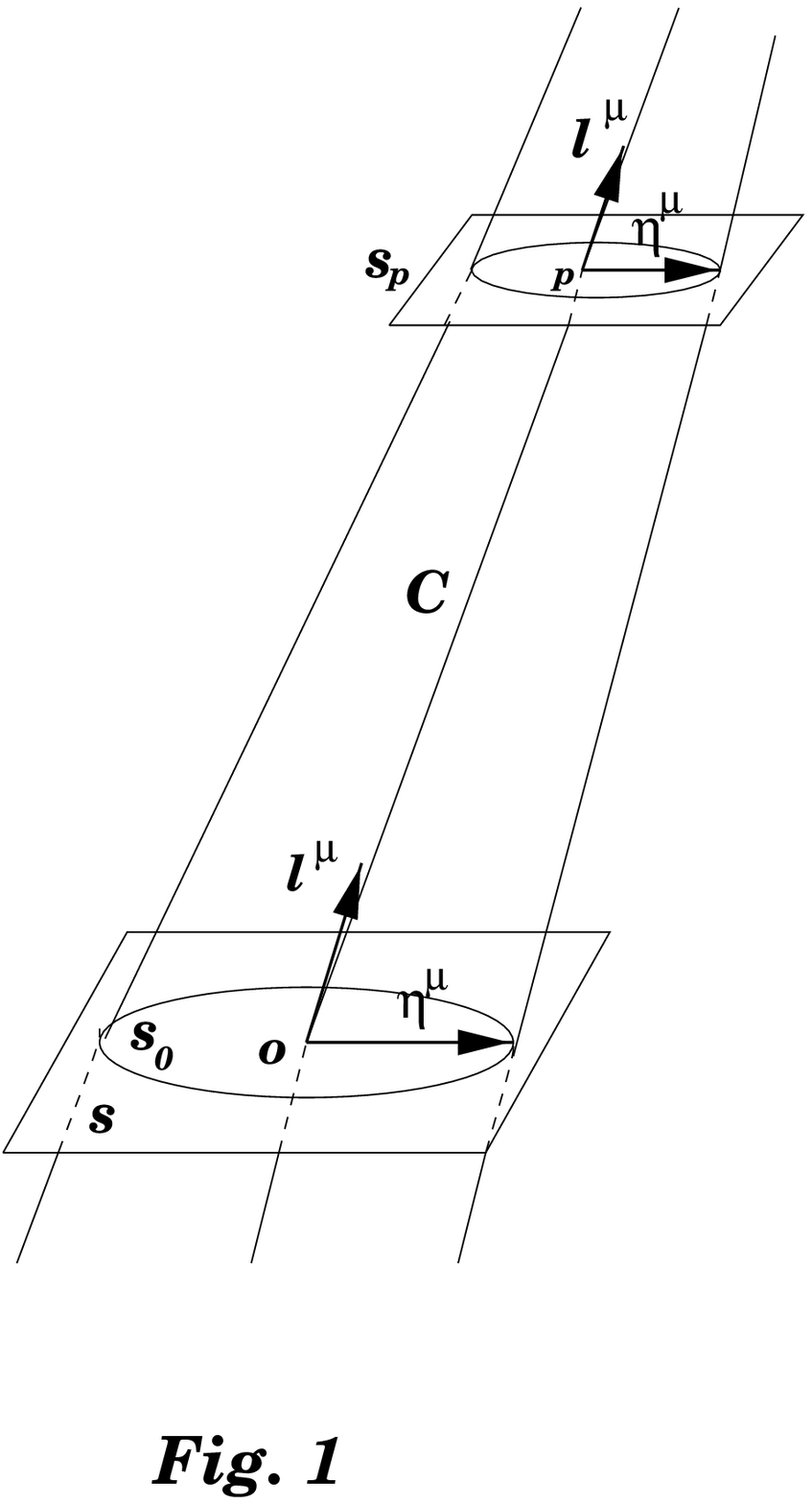}}

\newpage 

{\bf Fig.2} (a) The image of the circle $S$ on  $S_{P}$
is contracted because of the interaction of $\Phi_{00}$.
(b) The image of the circle $S$ on  $S_{P}$
is deflected into an ellipse with its main major axis along $e_{2}$
because of the interaction of $\Psi_{0}$.
%\vspace{1 cm}

\centerline{\epsffile{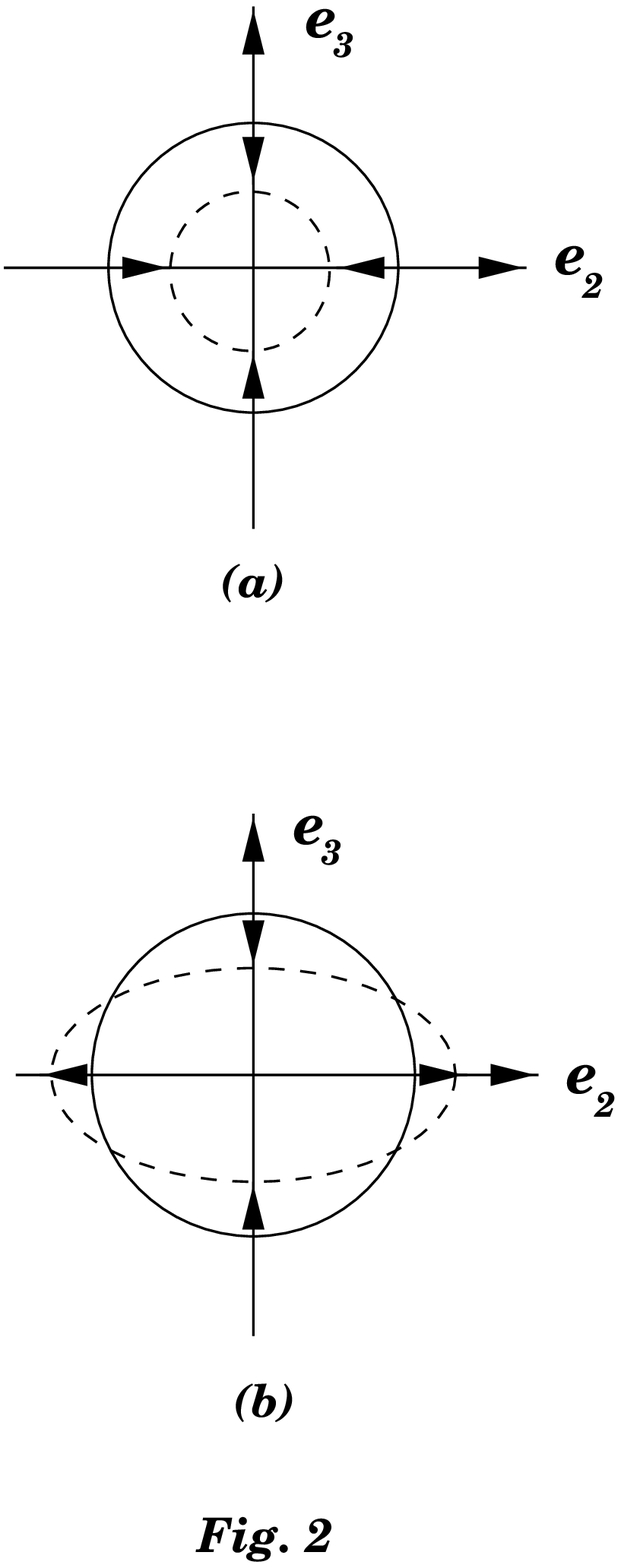}}

\newpage

{\bf Fig.3} A spherical ball consisting of photons
 cuts $S_{O}$ in the circle $S$ with the point $O$ as its center. The image
 of the ball at the point $P$ is turned into a
spheroid with the main major axis along a line at $45^{0}$ 
with respect to $e_{1}$ in the plane spanned by $e_{1}$ 
and $e_{3}$ because of the interaction of $\Psi_{1}$
and $\Phi_{01}$, while the rays are left undeflected  in the 
$e_{2}$-direction.

\vspace{1.5 cm}

\centerline{\epsffile{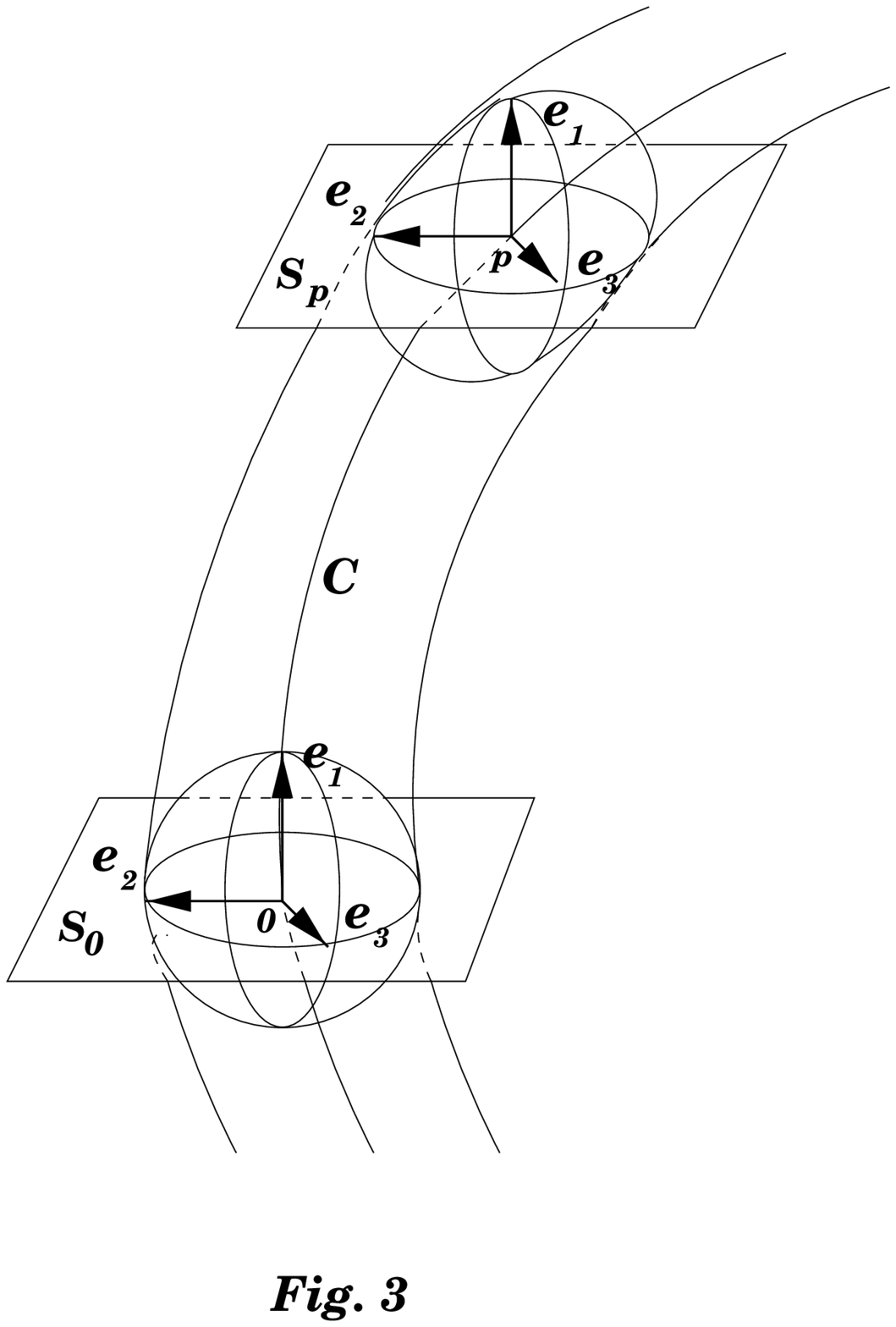}}

\end{document}